\documentclass[authoryear,preprint,review,12pt]{elsarticle}

\usepackage{amssymb}
\usepackage{graphicx}
\usepackage{color}
\newcommand{\logg}{\ensuremath{\log g}}                                 

\journal{New Astronomy}

\begin{document}

\begin{frontmatter}



\title{Complex Analysis of the Stellar Binary V446\,Cep; A New Massive 
Eclipsing Binary in Cepheus\,OB2 Association }

\author[l1]{\"{O}. \c{C}ak{\i}rl{\i} \corref{cer}}

\address[l1]{Ege University, Science Faculty, Astronomy and Space Sciences Dept., 35100 Bornova, \.{I}zmir, Turkey \corref{cer}}
\cortext[cer]{Corresponding author. Tel.: +902323111740; Fax: +902323731403 \\
E-mail address: omur.cakirli@gmail.com}

\author[l1]{C. Ibanoglu}
\author[l1]{E. Sipahi}
\author[l3]{A. Frasca}
\author[l3]{G. Catanzaro}
\address[l3]{Osservatorio Astrofisico di Catania, Via S. Sofia 78, 95123 Catania, Italy}

\begin{abstract}
We present new spectroscopic observations of the early type, double-lined eclipsing binary 
V446\,Cep. The radial velocities and the photometric data obtained by $Hipparcos$ were 
analysed for deriving the astrophysical parameters of the components. Masses and radii were determined as 
M$_p$=17.94$\pm$1.16 M$_{\odot}$ and R$_p$=8.33$\pm$0.29 R$_{\odot}$, M$_s$=2.64$\pm$0.30 M$_{\odot}$ and
R$_s$=2.13$\pm$0.10 R$_{\odot}$ for the components of V446\,Cep. Our analyses show that V446\,Cep is a 
detached Algol-type system. Based on the position of the components plotted on the theoretical 
Hertzsprung-Russell diagram, we estimate that the age of V446\,Cep is about 10 Myr, neglecting the 
effects of mass-loss and mass exchange between the components. Using the UBVJHK magnitudes and interstellar 
absorption we estimated the mean distance to the system V446\,Cep  as 1100$\pm$62\,pc. 
\end{abstract}

\begin{keyword}
stars: binaries: eclipsing -- stars: fundamental parameters -- stars: binaries: spectroscopic -- stars:V446\,Cep
\end{keyword}

\end{frontmatter}

\section{INTRODUCTION }
High-mass stars are much less frequent than intermediate- or low-mass stars due to both the star formation
process, which gives rise to an initial mass function  declining with the mass \citep[e.g.][]{Salpeter1955, Kroupa2001},	
and to their shorter evolutionary times. However, high-mass stars are very important because they can affect 
their surroundings with their winds, their strong radiation fields, and their catastrophic death as supernovae, 
chemically enriching their environment and triggering star formation. They usually form within the dense cores 
of stellar clusters and/or associations where dynamical interactions play an important role. It is widely believed 
that massive stars may be the product of collisions between two or more intermediate-mass stars. This idea is 
supported by the fact that a large fraction of massive stars harbour close companions, as failed mergers. It has 
been recently estimated that at least 50\,\% of massive stars are member of binary or multiple star system 
\citep{San12}. This lucky occurrence allows to directly measure the masses by means of their radial velocity 
(RV) curves. In many cases, spectral lines of both components are visible (SB2 systems), allowing to derive 
the orbital parameters like the period, $P_{\rm orb}$, the projected semi-major axes, $a_{1,2}\sin i$, and 
the masses, $M_{1,2}\sin^3i$, apart from the factor $\sin^3i$. If our line-of-sight is close to the orbital 
plane and fractional radii of the components are not too small the stars display mutually eclipses. The orbital 
inclination and fractional radii of the component stars can be determined by the analysis of photometric light 
curves. Therefore, eclipsing binaries are unique targets for determining the masses and radii from their 
combined light curves and radial velocities analyses. Nevertheless, absolute radii were measured only for 
a rather small number of early-type B-stars which are members of eclipsing binary systems \citep{Hil04, Tor10, 
Iba03, Iban13}. Thus, we started a systematic observing program devoted to the spectroscopic study of close 
eclipsing binary systems with at least one hot component.

\subsection{V446\,Cephei} 
The light variability of V446\,Cephei (HD 210478; BD+60$^{o}$2348; HIP 109311; V=7$^{m}$.32, B-V=0$^{m}$.08) has 
been discovered by the $Hipparcos$ photometry \citep{Per97}. It was classified as an Algol-type eclipsing 
binary (EA) from visual inspection of light variation.  The orbital 
period of V446\,Cep has been determined by the $Hipparcos$ satellite as 3.8084 days and the ephemeris is 
given as follows,
\begin{equation}
Min I(HJD)=2\,448\,503.047+3^d.8084 \times E.
\end{equation}
The $Hipparcos$' light curve (LC) displays a primary minimum with a depth of about 0.14 mag \citep{Mal06}. \citet{Kaz99} 
designated it as V446\,Cep and classified it as an EA according to the criteria of the General Catalogue of 
Variable Stars. \citet{Sim68} included the star in the list of probable members of the Cep\,OB2 association. 

In this paper we present new spectroscopic observations of V446\,Cep. We evaluate effective temperatures and surface 
gravities of the component stars from an ad-hoc analysis of the spectra taken near the quadratures. 
Combining the results obtained by the analysis of RV and light curves we determine directly absolute masses 
and radii of the components. We infer distance to the system, their fundamental parameters, and briefly 
discuss the main outcomes of this study.

\section{OBSERVATIONS}
Optical spectroscopic observations of V446\,Cep were obtained at the TUBITAK National Observatory using
the Turkish Faint Object Spectrograph Camera (TFOSC)
attached to the 1.5\,m telescope. The observations were made from July 22, 2012 to August 3, 2013, under good seeing
conditions. Further details on the telescope and the spectrograph can be found at http://www.tug.tubitak.gov.tr. The 
wavelength coverage of each spectrum was 4000-9000 \AA~in 11 orders, with a resolving power of 
$\lambda$/$\Delta \lambda$ $\sim$7\,000 at 6500 \AA. The average signal-to-noise ratio (S/N) was $\sim$120. We also 
obtained high S/N spectra of early type standard stars 1\,Cas (B0.5\,IV), HR\,153 (B2\,IV), $\tau$\,Her (B5\,IV), 
21\,Peg (B9.5\,V) and $\alpha$\,Lyr (A0\,V) which were used as templates in derivation of the radial velocities. 

The echelle spectra were extracted from the raw images following standard reduction steps involving electronic 
bias subtraction, flat field division, cosmic rays removal, optimal extraction of the echelle orders, and 
wavelength calibration thanks to the emission lines of a Th-Ar lamp. The reduction was performed using tasks 
of the IRAF package\footnote{IRAF is distributed by the National Optical Astronomy Observatory, which is 
operated by the Association of Universities for Research in Astronomy, Inc. (AURA), under cooperative 
agreement with the National Science Foundation.}

\section{RADIAL VELOCITIES AND ATMOSPHERIC PARAMETERS}
Our spectroscopic dataset contains 15 observations for V446\,Cep. We have measured radial 
velocities (RVs) from the spectra, focusing on spectral segments containing the He\,{\sc i} $\lambda$5876 
(order\,4) and $\lambda$6678 (order\,3) lines which are the most prominent un-blended features in our 
spectra, apart from the Balmer lines. We have employed the standard cross-correlation method for measuring 
the velocities of the component stars of the system. The cross-correlation technique \citep{Sim74,Ton79} 
is widely used for measuring RVs from the spectra of close binary systems. Cross-correlation analyses 
were made using the spectra of $\tau$\,Her and 21\,Peg as templates. The principle spectral features showing 
splitting due to binarity were the  He\,{\sc i} lines at $\lambda\lambda$5876 and 6678. We used also order 9, 
containing the He\,{\sc i} $\lambda$4471 line, for a few measurements of the radial velocities. The spectra taken 
close to the conjunctions, which display no double-lined feature, were disregarded. The Balmer lines were not used 
in the measurements of radial velocities due to their extremely broad profiles.

We obtained 15 radial velocities for each component of V446\,Cep. The average radial velocities and their associated 
standard errors derived from the spectral segments containing He\,{\sc i} $\lambda\lambda$4471, 5876, 
and 6678 lines are presented in Table\,1, along with the observation date and orbital phase. The mean error of 
radial velocities is 2.5\,km\,s$^{-1}$ for the primary, and 3.6 km\,s$^{-1}$ for the secondary star of V446\,Cep. The 
RVs are plotted against the orbital phase in Fig.\,1, where the filled squares represent the primary and the empty 
squares the secondary stars. Examination of the $Hipparcos$ light curve show no evidence for any eccentricity 
in the orbit of the system. Therefore, we have assumed circular orbit and analysed the RVs using the {\sc RVSIM} 
software programme \citep{Kan07}. Final orbital parameters are presented in Table\,2. 

\subsection{Determination of the atmospheric parameters}
Intermediate-resolution optical spectroscopy permits us to derive most of the fundamental stellar parameters, such as the
projected rotational velocity ($v\,sin\,i$), spectral type (S$_p$), luminosity class, effective temperature 
($T_{\rm eff}$), surface gravity ($log~g$), and metallicity ([Fe/H]).

The width of the cross-correlation function (CCF) is a good tool for the measurement of projected rotational 
velocity ($v\sin i$) of a star. We use a method developed by \citet{Pen96} to estimate the $v\sin i$ of each 
star composing the investigated SB2 system from its CCF peak by a proper calibration based on a spectrum of 
a narrow-lined star with a similar spectral type. The rotational velocities of the components 
were obtained by measuring the FWHM of the CCF peak related to each component in five high-S/N spectra 
acquired near the quadratures, where the spectral lines have the largest Doppler-shift. The CCFs were 
used for the determination of $v\,sin\,i$ through a calibration of the full-width at half maximum (FWHM) of 
the CCF peak as a function of the $v\,sin\,i$ of artificially broadened spectra of slowly rotating standard 
star (21\,Peg, $v\sin i \simeq$14 km\,s$^{-1}$, e.g., \citealt{Roy02}) acquired with the same setup and in 
the same observing night as the target system. The limb darkening coefficient was fixed at the theoretically
predicted values, 0.42 for the system \citep{Van93}.  We calibrated the relationship between the CCF Gaussian
width and $v\,sin\,i$ using the \citet{Con77} data sample. This analysis yielded projected rotational
velocities for the components of V446\,Cep as $v_{\rm P}\sin i$=120~km\,s$^{-1}$, and $v_{\rm S}\sin i=44$~km\,s$^{-1}$. 
The mean deviations were 3 and 9 km\,s$^{-1}$, for the primary and secondary, respectively, between the measured
velocities for different lines.

We also performed a spectral classification for the components of the system using COMPO2, an IDL code for the analysis 
of high-resolution spectra of SB2 systems written by one of us \citep[see, e.g.,][]{Frasca2006} and adapted to the TFOSC 
spectra. This code searches for the best combination of two reference spectra able to reproduce the observed spectrum of 
the system. We give, as input parameters, the radial velocities and projected rotational velocities $v\sin i$ of the two 
components of the system, which were already derived. The code then finds, for the selected spectral region, the spectral 
types and fractional flux contributions that better reproduce the observed spectrum, i.e. which minimize the residuals in 
the collection of difference (observed\,$-$\,composite) spectra. For this task we used reference spectra taken from the 
\citet{Val04} $Indo-U.S.\ Library\ of\ Coude\ Feed\ Stellar\ Spectra$ (with a a resolution of $\approx$\,1\AA) that are 
representative of stars with spectral types from late-O type to early-A, and luminosity classes V, IV, and III. 
The atmospheric parameters of these reference stars were recently revised by \citet{Wu2011}.

We selected 198 reference spectra spanning the ranges of expected atmospheric parameters, which means that we have searched 
for the best combination of spectra among 39204 possibilities per each spectrum. The observed spectra of V446\,Cep in the 
$\lambda\lambda$6525--6720 spectral region were best represented by the combination of HD\,187459 (B0.5 II) and HD\,178125 
(B8 III). However, we have adopted, for each component, the spectral type and luminosity class with the highest score in the 
collection of the best combinations of templates, where the score takes into account the goodness of the fit expressed by 
the minimum of the residuals. We have thus derived a spectral types for the primary and secondary component of V446\,Cep as
B1 and B9 main-sequence stars, with an uncertainty of about 1 spectral subclass.
The effective temperature and surface gravity 
of the two components of the system are obtained as the weighted average of the values of the best spectra at phases near to the 
quadratures combinations of templates adopting a weight $w_i=1/\sigma_i^2$, where $\sigma_i$ is the average of residuals 
for the $i$-th combination. The standard error of the weighted mean was adopted for the atmospheric parameters. Both stars 
appear to have a solar metallicity, within the errors. The atmospheric parameters obtained by the code and their standard 
errors are reported in Table\,3. The observed spectra of V446\,Cep at phases near to the quadratures are shown in Fig.\,2 
together with the combination of two reference spectra which gives the best match.

\section{LIGHT CURVE ANALYSES}
The light curve of V446\,Cep was obtained by the $Hipparcos$ spacecraft and is composed of 124 photometric points. The light
curve is very similar to that of a detached Algol-type binary. The brightness of the system at the maximum and depth of 
the primary minimum were estimated by \citet{Mal06} to be 7.31 and 0.14 mag, respectively. However, we estimate the brightness of 
the system at the maximum light as Hp=7.348 mag, with a mean error of about 0.013 mag, and the depth the primary minimum 
as 0.10 mag. The Hp magnitudes were transformed to the Johnson's V-passband using the coefficients given by \citet{Harm98}. 
The $Hipparcos$ light curve is plotted against the orbital phase in Fig.\,3. There is no indications of any asymmetry in the LC. 
As the secondary minimum occurs at phase 0.5, we have adopted circular orbits for our analysis. The effective temperatures 
for the primary and secondary star were estimated from the spectra as 26\,580$\pm$880 and 12\,000$\pm$1050 K, respectively. 

The apparent visual magnitude and colour indices were given by \citet{Ree03} as $V$=7$^m$.32, ($U-B$)=$-$0$^m$.71, and 
($B-V$)=0$^m$.08. We obtained the reddening-free quantity  $Q$=$-$0.768, which corresponds to a B1 main-sequence star 
\citep{Hov04} with an intrinsic colour of ($B-V$)=$-$0$^m$.273. This colour index corresponds to an effective temperature 
of 26\,800$\pm$800 K \citep{Flo96}, which is consistent with that obtained directly from the spectra. A preliminary 
analysis of the light curve gives a light ratio of $l_{s}$/$l_{p}$=0.014. Using the intrinsic colour of the primary 
star, the light ratio and the observed composite colour of the system an interstellar reddening of $E_{(B-V)}$=0$^m$.338 
was determined for the system.    

We started to analyze the light curve using the Wilson-Devinney code \citep[hereafter WD; e.g.,][]{Wil71,Wil79,Wil06}
as implemented in the software {\sc phoebe} \citep{Prs05}. The WD code is widely used for determination of the orbital 
parameters of the eclipsing binaries. To run the code we need some initial parameters. The initial logarithmic 
limb-darkening coefficients were taken from the tables given by \citet{Van93} as $x_1$=0.41 and $x_2$=0.62, $y_1$=0.24 
$y_1$=0.30 and are automatically interpolated at each iteration by {\sc phoebe}. The effective temperature of the 
primary star is taken as 26\,600 K and the ratio of the masses of two components $q$=0.147. We have started with 
$Mode-2$ meant for detached binary systems, keeping the temperature of the primary and the mass-ratio as fixed 
parameters. According to the WD code we  adjusted the following parameters: $i$ (the orbital inclination), 
$\Omega_1$ (the potential for the primary), $\Omega_2$ (the potential for the secondary), $T_{\rm eff_2}$ (the 
effective temperature of the cool star), L$_1$ (the luminosity of the primary), and the zero-epoch offset. The 
luminosity of the secondary star, L$_2$,  was constrained by the model. After a few numbers of runs of the DC 
program in $Mode-2$ the sum of residuals squared showed a minimum and the corrections to the adjustable parameters 
became smaller than their probable errors. The results are given in the Table\,4. The corresponding 
computed light curve is shown in Fig.\,3 as a continuous line.

\section{Results and discussion}   
Based on the results of radial velocities and light curves analyses we have calculated the physical properties 
of the V446\,Cep. For this purpose, we used the $JKTABSDIM$ code developed by \citet{Sou05}. 
This code is now widely used for derivation of the absolute parameters of the eclipsing binary stars' components.
It calculates complete error budgets using a perturbation algorithm. The fundamental stellar parameters for the
components such as masses, radii, luminosities and their standard deviations have been derived using this code. 
The astrophysical parameters of the components, and other properties for the stars of V446\,Cep is presented
in Table\,5.

The separation between the components was found to be 28.10$\pm$0.63 R$_{\odot}$ for V446\,Cep. The masses were 
measured to precision of about 6--7\,\%, apart from the mass of the secondary star of V446\,Cep, which has an 
uncertainty of about 11\,\%. On the other hand the radii of the stars have been derived with a precision of better 
than 5\,\%. The accuracy of any parameter of an eclipsing binary system depends mainly on the coverage of the both 
spectroscopic and photometric observations and their precision. In addition, the LC solutions are more accurate 
for totally eclipsing systems. The light curve of V446\,Cep shows total eclipses, but the precision of the photometric 
measurements is not sufficiently high. Despite these drawbacks, the physical parameters of the components of system 
could be determined with sufficient precision. We note that the effective temperature of the secondary star derived 
from the spectra is in good agreement with that obtained from the light curve analysis.

The luminosities and absolute bolometric magnitudes are calculated directly from the radii and and effective
temperatures of the components. The effective temperature of 5\,777~K and the absolute bolometric magnitude of
4.74 mag were adopted for the Sun \citep[e.g.,][]{Dri00}. The bolometric corrections were interpolated from the
tables of \citet{Flo96}. The V-band magnitude of the system at out-of-eclipse phases is taken as 7.32
for V446\,Cep. We have calculated the absolute visual magnitudes for the components using the fractional luminosities 
and bolometric corrections given in Table\,4 and 5. Combining these values with the interstellar absorption of 1.09\,mag 
for V446\,Cep we have estimated the distance to the system as 1100$\pm$62\,pc.

In the log $T_{\rm eff}$--$\log L/L_{\odot}$ (left panel) and log $T_{\rm eff}$-log\,g planes (right panel) of Fig.\,4 we have 
plotted the positions of the components, with 1-$\sigma$ error bars. The filled and empty circles represent 
the primary and secondary star of V446\,Cep. The evolutionary tracks and isochrones for the non-rotating single stars 
with solar composition are taken from \citet{Eks12}.

V446\,Cep was included in the list of probable members of the Cep\,OB2 association \citep[e.g.,][and reference therein]{Sim68}.
According to \citet{Pat98} the distance to the Cepheus OB2 association is $\sim$\,900\,pc and it is embedded in the 
Cepheus bubble, a giant shell structure of atomic and molecular gas extending in a radius of about 120\,pc, which is
believed to be generated by an earlier generation of hot and massive stars in NGC\,7160. \citet{Gar92} derived
a reddening in the range 0.29--1.12 for the stars in the Cep\,OB2 field, with a mean value of 0.59 for the members of 
the association. The interstellar reddening that we estimated for V446\,Cep, $E_{(B-V)}$=0.34, is lower than the value 
estimated for the association, but it is consistent with the value of 0.32 derived by \citet{Har98} for the binary 
system LZ\,Cep in this association. 

This result suggests that the binary is located near the NGC\,7160  in the border of Cepheus Bubble \citep{Kun08}.
The open clusters Tr\,37 and NGC\,7160 with the ages of 4 and 12 Myr locate in the association \citep{Sic04}.
Our distance of V446\,Cep agrees quite well with those distance values of Cep\,OB2 association. Both components of
V446\,Cep appear to have an age of about 10 Myr which is consistent with the age of stars in the open cluster NGC\,7160.
 
As seen from Fig.\,4, components of the system are located on the main-sequence, i.e. in the central hydrogen burning phase. They are 
still inside their lobes, corresponding to the detached Algols. \citet{Iba06} collected the physical parameters of 74 
detached Algols, mainly composed of hot (BAF-type) stars. Their mass-ratios, M$_{2}$/M$_{1}$, are generally larger 
than 0.5, with a mean value of 0.88. The binary system  AR\,Cas has the smallest mass-ratio in their Table\,1, with 
a value of 0.315. We find V446\,Cep as a detached Algol with a lower mass-ratio. $M_2/M_1=0.147$, which is among 
lowest values found for this class of binaries. The less massive stars in the semi-detached binaries fill their 
corresponding Roche lobes and are oversized and over-luminous relative to a zero-age main-sequence star of the 
same mass \citep{Har98}. Although the less-massive secondary star of V446\,Cep does not fill its respective Roche 
volume it is seen as oversized and over-luminous with respect to its mass, similar to the secondaries of the 
semi-detached systems.

\section{Summary} 
V446\,Cep is close eclipsing binary containing high-mass star. We carried out spectroscopic observations
of the system. The atmospheric parameters of the stars in the eclipsing pair have been determined from its 
spectra. The spectra were analyzed using cross-correlation for measuring the radial velocities of both components 
and with an ad-hoc code for deriving their atmospheric parameters. Moreover, $HIPPARCOS$ light curve was modeled 
using the WD code. The physical parameters for the system are measured to accuracies of 6-7\,\% in mass, apart 
from the secondary of V446\,Cep, and 5\,\% in radius. The distance to the system V446\,Cep was estimated as 
1100$\pm$62\,pc. A comparison of physical parameters of the components with the theoretical models 
of single stellar evolution models has been made and an age of about 10\,Myr for V446\,Cep has been derived. Both the estimated age and 
distance of V446\,Cep confirm its membership of Cepheus OB2 association, locating close to the open cluster NGC\,7160.

\section*{Acknowledgments}
We thank to T\"{U}B{\.I}TAK National Observatory (TUG) for a partial support in using RTT150 
telescope with project number {\sf 11BRTT150-198}.
We thank to EB{\.I}LTEM Ege University Research Center for a partial support with project number {\sf 2013/BIL/018}.
We also thank to the staff of the Bak{\i}rl{\i}tepe observing station for their warm hospitality. This study is 
supported by Turkish Scientific and Technology Council under project number {\sf 112T263}.
This research was also partly supported by the Scientific Research Projects Coordination Unit of
Istanbul University. Project number 3685. We thank Canakkale Onsekiz Mart University Astrophysics
Research Center and Ulupinar Observatory together with Istanbul University Observatory
Research and Application Center for their support and allowing use of IST60 telescope.
This work was partially supported by the Italian {\em Ministero dell'Istruzione, Universit\`a e  Ricerca} (MIUR).
The following internet-based resources were used in research for this paper: the NASA Astrophysics Data System; the 
SIMBAD database operated at CDS, Strasbourg, France; and the ar$\chi$iv scientific paper preprint 
service operated by Cornell University. 

\newpage
\begin{figure}
\center
\hspace{-1.2cm}
\includegraphics[width=9.3cm,angle=0]{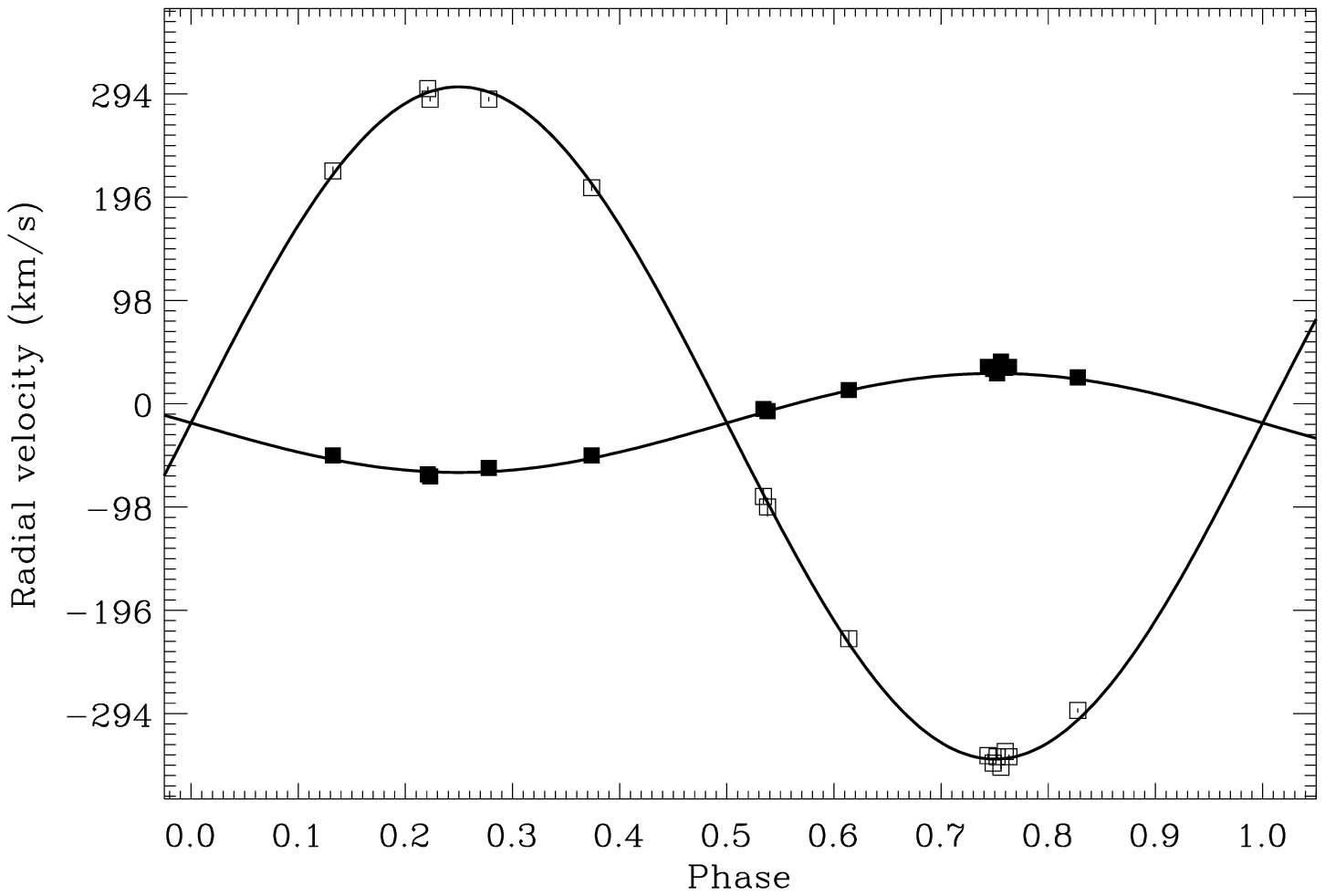}
\caption{Radial velocities for the components of V446\,Cep. Filled squares correspond to the radial velocities 
for the primary and the empty squares for the secondary star. Error bars are shown by vertical line segments, 
which are smaller than symbol sizes. The solid lines are the computed radial velocity curves for the component stars.} 
\end{figure}

\begin{figure*}
  \begin{center}
  \includegraphics[width=10cm]{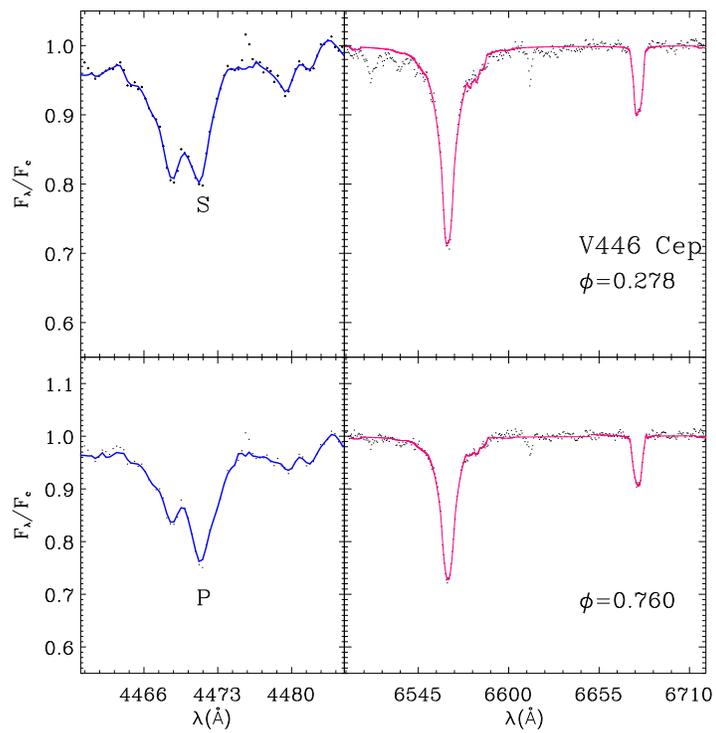}
  \end{center}
  \caption{ Comparison between the observed spectra of V446\,Cep and the best-fitting spectra around  He\,{\sc i} $\lambda$4471 (left panel) and  the H$\alpha$ and He\,{\sc i} $\lambda$6678 lines (right panel).}
\end{figure*}

\begin{figure}
  \begin{center}
  \includegraphics[width=8.5cm]{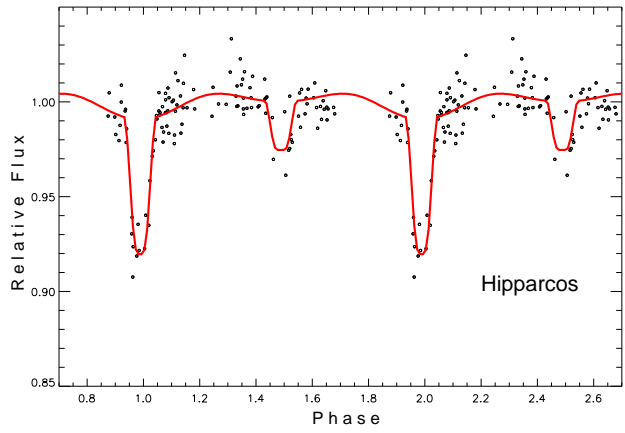}
  \end{center}
  \caption{The $Hipparcos$ light curve of V446\,Cep. The continuous line shows the best-fit model.}  \label{fig:evo}
\end{figure}

\begin{figure*}
  \begin{center}
 \includegraphics[width=14cm]{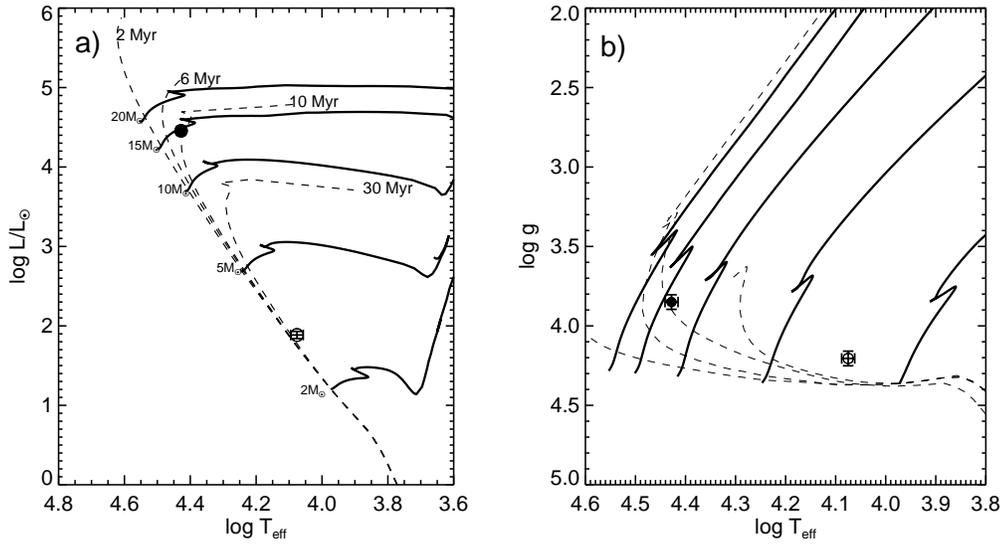}
  \end{center}
  \caption{Positions of the components of the system in the luminosity-effective temperature and 
gravity effective temperature planes are plotted. The filled and empty circles represent the primary 
and secondary stars, respectively, of V446\,Cep. The solid lines show evolutionary tracks for single 
stars with masses of 20, 15, 10, 5 and 2 solar masses for solar composition taken from 
\citet{Eks12}. The positions of the components of V446\,Cep are consistent with an age of $\sim$ 
10 Myr. }  \label{fig:evo}
\end{figure*}

\begin{table}
\centering
\begin{minipage}{85mm}
\caption{Heliocentric radial velocities of V446\,Cep. The columns give the heliocentric 
Julian date, orbital phase and the radial velocities of the two components with the corresponding 
standard deviations.}
\begin{tabular}{ccrcrcc}
\hline
HJD 2400000+ & Phase & \multicolumn{4}{c}{V446\,Cep }&  	\\
             &       & $V_{\rm P}$                      & $\sigma$                    & $V_{\rm S}$  	& $\sigma$	\\
\hline
56132.3303	&	0.2778	&	$-61$	&	2	&	$ 289$    &	  2	  \\
56134.4239	&	0.8276	&	$ 25$	&	2	&	$-291$    &	  2	  \\
56135.5843	&	0.1323	&	$-49$	&	3	&	$ 221$    &	  4	  \\
56136.5042	&	0.3738	&	$-49$	&	3	&	$ 205$    &	  3	  \\
56137.4177	&	0.6137	&	$ 13$	&	3	&	$-223$    &	  8	  \\
56162.5807	&	0.2209	&	$-67$	&	2	&	$ 299$    &	  2	  \\
56162.5890	&	0.2231	&	$-69$	&	2	&	$ 289$    &	  2	  \\
56168.3798	&	0.7436	&	$ 35$	&	2	&	$-334$    &	  2	  \\
56506.5293	&	0.5341	&	$ -5$	&	4	&	$ -88$    &	  7	  \\
56506.5444	&	0.5380	&	$ -7$	&	4	&	$ -98$    &	  9	  \\
56507.3464	&	0.7486	&	$ 33$	&	2	&	$-341$    &	  3	  \\
56507.3596	&	0.7521	&	$ 29$	&	2	&	$-335$    &	  3	  \\
56507.3739	&	0.7558	&	$ 40$	&	2	&	$-345$    &	  2	  \\
56507.3895	&	0.7599	&	$ 34$	&	2	&	$-330$    &	  3	  \\
56507.4025	&	0.7633	&	$ 35$	&	2	&	$-335$    &	  2	  \\
\hline
\end{tabular}
\end{minipage}
\end{table}

\begin{table}
\centering
\begin{minipage}{85mm}
\caption {Results of the radial velocity analysis for V446\,Cep.}
\begin{tabular}{@{}ccccccccc@{}c}
\hline
Parameter  & \multicolumn{2}{c}{V446\,Cep }&  	\\
             & Primary           & Secondary   	\\
\hline
 $K$ (km\,s$^{-1}$) 			&$ 47\pm$4	      			&$319\pm$7	\\
    $V_\gamma$ (km\,s$^{-1}$) 		&\multicolumn{2}{c}{$-18\pm$3}  \\
    Average O-C (km\,s$^{-1}$)		& 2.5         	      			&3.6      	\\
    $a\sin i$ ($R_{\odot}$)		& 3.51$\pm$0.27	      			& 24.03$\pm$0.49\\ 
    $M\sin^3i$ ($M_{\odot}$)		& 16.86$\pm$1.08      			& 2.48$\pm$0.28	\\      
\hline
\end{tabular}
\end{minipage}
\end{table}

\begin{table}
\centering
\begin{minipage}{85mm}
\caption {Spectral types, effective temperatures, surface gravities, and rotational velocities of components 
estimated from the spectra of V446\,Cep.}
\begin{tabular}{@{}ccccccccc@{}c}
\hline
Parameter  & \multicolumn{2}{c}{\sf V446\,Cep }&  	\\
             & Primary           & Secondary   	\\
\hline
Spectral type 				& B(1$\pm$0.5)\,V	&B(9$\pm$0.5)\,V	\\
 $T_{\rm eff}$ (K)	    		&26\,580$\pm$880  	&12\,000$\pm$1050	\\   
 $\log~g$ ($cgs$)			& 3.77$\pm$0.05        	&3.95$\pm$0.17       	\\    
 $v\sin i$ (km\,s$^{-1}$)  		&120$\pm$3	  	& 44$\pm$9     		\\   
\hline
\end{tabular}
\end{minipage}
\end{table}

\begin{table}
\centering
\begin{minipage}{85mm}
\caption{Final solution parameters for the semi-detached model of V446\,Cep. }
\begin{tabular}{@{}ccccc}
\hline
Parameters  &  V446\,Cep 	\\                  
\hline
$i^{o}$			                    	       &78.36$\pm$0.22  	  \\
$T_{\rm eff_1}$ (K) 				       &26\,600[Fix]		  \\
$T_{\rm eff_2}$ (K) 				       &11\,900$\pm$1\,500	 \\
$\Omega_1$ 	          			       &3.452$\pm$0.132 	 \\
$\Omega_2$					       &3.217$\pm$0.119 		       \\
$r_1$				                       &0.2962$\pm$0.0118	 \\
$r_2$		           	            	       &0.0756$\pm$0.0046	 \\
$\frac{L_{1}}{(L_{1}+L_{2})}$   		       &0.9858$\pm$0.0036	 \\
$\sum(O-C)^{2}$			            	       &0.0108  		 \\	 
$N$		           	              	       &124			   \\	 
$\sigma$			                       &0.0097  		 \\				 
\hline  
\end{tabular}
\end{minipage}
\end{table}

\setlength{\tabcolsep}{3pt}
\begin{table}
\footnotesize
\centering
\begin{minipage}{85mm}
\caption{Absolute parameters, magnitudes and colours for the components of V446\,Cep. }
\begin{tabular}{@{}llcccccccc@{}}
\hline
Parameter & Units &\multicolumn{2}{c}{\sf V446\,Cep } 	\\
          &    & Primary                      & Secondary           \\
\hline
Mass                & M$_{\odot}$     & 17.94$\pm$1.16      &2.64$\pm$0.30	 \\ 
Radius              & R$_{\odot}$     & 8.33$\pm$0.29	    & 2.13$\pm$0.10	 \\
$T_{\rm eff}$       & K	              & 26\,600$\pm$1000    & 11\,900$\pm$1050   \\
$\log(L/L_{\odot})$ &	   	      & 4.500$\pm$0.056     & 1.904$\pm$0.118	 \\ 
$\logg$             & $cgs$	      & 3.850$\pm$0.023     & 4.205$\pm$0.050	 \\
$Sp.Type$           &		      & B1V		    & B8IV		 \\
$M_{bol}$           & mag             & $-6.51\pm$0.14      & $-0.02\pm$0.30	 \\
$BC$                & mag	      & $-2.58$ 	    & $-0.66$		 \\
$M_{V}$             & mag             & $-3.93\pm$0.08      & 0.64$\pm$0.17	 \\
$v\sin i_{\rm cal}$ & km\,s$^{-1}$    & 108$\pm$4	    & 28$\pm$1  	 \\
$v\sin i_{\rm obs}$ & km\,s$^{-1}$    & 120$\pm$8	    & 44$\pm$7  	 \\
$d$                 & pc              & 1\,100$\pm$62	    & 1\,120$\pm$40	 \\
\hline
\end{tabular}
\end{minipage}
\end{table}

\end{document}